\documentclass[a4paper,11pt]{article}
\usepackage[DIV10]{typearea}
\usepackage[T1]{fontenc}

\usepackage[bookmarksopen,colorlinks]{hyperref}
\usepackage{graphicx}
\usepackage[english]{babel}
\usepackage{amssymb}
\usepackage{geometry}
\usepackage{color}
\usepackage{xspace}
\usepackage{textcomp}
\usepackage{epic}
\usepackage{slashed}
\usepackage{eepic}
\usepackage{amsmath,bm}
\usepackage{epstopdf}
\usepackage{ifthen}
\usepackage{wasysym}
\usepackage[footnotesize]{caption2}
\hypersetup{
   pdfstartview={FitH},
   pdfpagemode={UseOutlines}
}

\DeclareMathSymbol{\Gamma}{\mathalpha}{letters}{"00}
\DeclareMathSymbol{\Delta}{\mathalpha}{letters}{"01}
\DeclareMathSymbol{\Theta}{\mathalpha}{letters}{"02}
\DeclareMathSymbol{\Lambda}{\mathalpha}{letters}{"03}
\DeclareMathSymbol{\Xi}{\mathalpha}{letters}{"04}
\DeclareMathSymbol{\Pi}{\mathalpha}{letters}{"05}
\DeclareMathSymbol{\Sigma}{\mathalpha}{letters}{"06}
\DeclareMathSymbol{\Upsilon}{\mathalpha}{letters}{"07}
\DeclareMathSymbol{\Phi}{\mathalpha}{letters}{"08}
\DeclareMathSymbol{\Psi}{\mathalpha}{letters}{"09}
\DeclareMathSymbol{\Omega}{\mathalpha}{letters}{"0A}
\DeclareMathSymbol{\varGamma}{\mathalpha}{operators}{"00}
\DeclareMathSymbol{\varDelta}{\mathalpha}{operators}{"01}
\DeclareMathSymbol{\varTheta}{\mathalpha}{operators}{"02}
\DeclareMathSymbol{\varLambda}{\mathalpha}{operators}{"03}
\DeclareMathSymbol{\varXi}{\mathalpha}{operators}{"04}
\DeclareMathSymbol{\varPi}{\mathalpha}{operators}{"05}
\DeclareMathSymbol{\varSigma}{\mathalpha}{operators}{"06}
\DeclareMathSymbol{\varUpsilon}{\mathalpha}{operators}{"07}
\DeclareMathSymbol{\varPhi}{\mathalpha}{operators}{"08}
\DeclareMathSymbol{\varPsi}{\mathalpha}{operators}{"09}
\DeclareMathSymbol{\varOmega}{\mathalpha}{operators}{"0A}

\newcommand{\D}{\mathrm{d}}

\def\beq{\begin{equation}}
\def\eeq{\end{equation}}
\def\bea{\begin{eqnarray}}
\def\eea{\end{eqnarray}}
\def\bi{\begin{itemize}}
\def\ei{\end{itemize}}

\newcommand{\stau}{{\widetilde{\tau}}}

\newcommand{\mstau}{m_{\stau_1}}

\newcommand{\mne}{{m_{\s{\chi}^0_1}}}
  
\newcommand{\msq}{{m_{\widetilde{q}}}}
\newcommand{\mgo}{{m_{\widetilde{g}}}}

\newcommand{\LCDM}{$\mathrm\Lambda\text{CDM}$}
\newcommand{\mG}{m_{\s G}}
\newcommand{\xf}{x_{\text{f}}}
\newcommand{\TR}{T_{\text{R}}}

\newcommand{\thest}{\theta_{\stau}}

\newcommand{\GEV}{\ensuremath{\,\textnormal{GeV}}}
\newcommand{\TEV}{\ensuremath{\,\textnormal{TeV}}}
\newcommand{\SEC}{\ensuremath{\,\textnormal{sec}}}

\newcommand{\fb}{\ensuremath{\,\textnormal{fb}}}

\newcommand{\s}[1]{\widetilde{#1}}

\newcommand{\HB}{{\textsc{HiggsBounds}}}

\newcommand{\FH}{{\textsc{FeynHiggs}}}

\newcommand{\sveff}{\langle \sigma_\text{eff}\,v_{\text{M\o l}} \rangle}

\newcommand{\Tf}{T_{\text{f}}}

\newcommand{\Mp}{M_{\text{Pl}}}

\newcommand{\G}{\s G}

\begin{document}

\date{\mbox{ }}

\title{ 
{\normalsize  
October $23^{\text{rd}}$, 2013 \hfill\mbox{}\\}
\vspace{2cm}
\bf
Gravitino LSP and leptogenesis after the first LHC results
\\[8mm]}

\author{Jan Heisig  \\[2mm] 
{\small
{\it Institute for Theoretical Particle Physics and Cosmology, RWTH Aachen, Germany}}
\\
{\small\tt heisig@physik.rwth-aachen.de}
}

\maketitle

\thispagestyle{empty}

\vspace{1cm}

\begin{abstract}

Supersymmetric scenarios where the lightest superparticle (LSP) is the gravitino 
are an attractive alternative to the widely studied case of a neutralino LSP\@. 
A strong motivation for a gravitino LSP arises from the possibility of achieving higher 
reheating temperatures and thus potentially allow for thermal leptogenesis.
The predictions for the primordial abundances of light elements in the presence of 
a late decaying next-to-LSP (NSLP) as well as the currently measured dark matter 
abundance allow us to probe the cosmological viability of such a scenario. 
Here we consider a gravitino-stau scenario. Utilizing a pMSSM scan we work out 
the implications of the 7 and 8\,TeV LHC results as well as other experimental and 
theoretical constraints on the highest reheating temperatures that are cosmologically 
allowed. Our analysis shows that points with $\TR\gtrsim10^9\GEV$ survive only in a 
very particular corner of the SUSY parameter space. Those spectra feature a distinct 
signature at colliders that could be looked at in the upcoming LHC run.

\end{abstract}

\clearpage

\section{Introduction}

The phenomenology of supersymmetric scenarios both at colliders and in the early
universe depends strongly on the nature of the lightest supersymmetric 
particle (LSP). The LSP is stable in the $R$-parity conserving case and thus
is usually identified with the dark matter (DM) candidate, if supersymmetry (SUSY) 
is to explain this observation. In a neutralino LSP scenario with a gravitino mass
of the order of the other sparticle masses, a cosmological problem appears
once we want to explain the observed baryon asymmetry in the universe
with the mechanism of thermal leptogenesis \cite{Fukugita:1986hr}. 
For this mechanism to work the universe has to be heated up to temperatures 
of $\TR\gtrsim10^{9}\GEV$ \cite{Davidson:2002qv,Buchmuller:2004nz} in the 
post-inflationary phase of reheating. On the other hand, from thermal scattering 
in the hot plasma gravitinos are produced \cite{Ellis:1984eq,Moroi:1993mb} and the
abundance of thermally produced gravitinos is proportional to $\TR$ 
\cite{Bolz:1998ek,Bolz:2000fu,Pradler:2006hh}.
Hence, a large $\TR$ leads to a large number density of gravitinos in the
early universe. The Planck-suppressed couplings of the gravitino lead to
a delayed decay into the LSP\@. These decays cause an 
additional energy release at or after the time of big bang nucleosynthesis 
(BBN) \cite{Ellis:1984eq,Falomkin:1984eu,Ellis:1984er}. 
The abundances of light elements are very sensitive to such 
processes and thus from their precise determination strong bounds can 
be imposed on the abundance of late-decaying gravitinos \cite{Kawasaki:1994af}. 
These bounds clearly exclude a reheating temperature of $\TR\gtrsim 10^9\GEV$. 
This problem is known as the gravitino problem \cite{Weinberg:1982zq}.

One way of avoiding this problem is a gravitino LSP scenario. Indeed, the gravitino
is a perfectly good DM candidate \cite{Fayet:1981sq,Pagels:1981ke}. 
However, in this scenario the next-to-LSP (NLSP) usually becomes long-lived 
and might spoil successful BBN predictions~\cite{Moroi:1993mb}. In contrast to the 
former scenario, it is now the abundance (and the life-time) of the late-decaying
NLSP which governs the phenomenological viability of the scenario in this concern. 
For an NLSP belonging to the sparticles of the MSSM---sharing the SM interactions---the
abundance is determined by the thermal freeze-out (rather than the reheating temperature). 
The abundance of the NLSP depends upon the spectrum parameters of the model and could, 
in principle, be determined from measurements at colliders. One of the most promising NLSP 
candidates in this concern is a charged slepton leading to a rather clean signature at colliders 
\cite{Fairbairn:2006gg,Raklev:2009mg}. In the upcoming high-energy run of the LHC such a 
scenario could reveal a rich phenomenology.

In a gravitino LSP scenario the maximally allowed reheating temperature can be constrained
from the measured DM abundance. Since the abundance of thermally produced gravitinos is 
approximately inversely proportional to the gravitino mass \cite{Bolz:1998ek,Bolz:2000fu,
Pradler:2006hh}, heavy gravitinos are favored from the requirement of large reheating 
temperatures whilst avoiding an over-closure of the universe.
On the other hand, the gravitino mass governs the life-time of the NLSP\@. Since BBN bounds 
disfavor extremely large life-times, those bounds become more constraining for larger gravitino 
masses. This non-trivial interplay can be used to formulate upper bounds on the reheating 
temperature \cite{Moroi:1993mb,Pradler:2006hh,Asaka:2000zh,Fujii:2003nr,
Roszkowski:2004jd,Pradler:2006qh,Steffen:2008bt,Choi:2008qh,Olechowski:2009bd,
Endo:2010ya,Endo:2011uw} on different levels of underlying assumptions.

In this work we consider a gravitino-stau scenario. 
We do not restrict ourselves to any constrained high-scale model but vary the SUSY parameters 
freely at the TeV-scale in the framework of the phenomenological Minimal Supersymmetric SM 
(pMSSM)~\cite{Djouadi:1998di}. Thereby we relax the particularly constraining
\cite{Fujii:2003nr,Pradler:2006qh,Steffen:2008bt} assumption of universal gaugino masses.
Further, in this study we include the non-thermal production of gravitinos through the decay 
of the stau NLSP\@.
This contribution can be very important for small mass differences between the
stau and the gravitino and introduces a further dependence of the allowed values for the reheating 
temperature on the SUSY parameters. Consequently, low stau abundances are favored in two
ways: by BBN constraints \emph{and} by the desire for a small contribution of non-thermal gravitino 
production.

In \cite{Heisig:2013rya} a survey for low stau abundances was performed in a Monte Carlo scan 
over a 17-dimensional pMSSM parameter space. In particular, the implications of a Higgs of around 
$125\GEV$, constraints from direct SUSY searches, from MSSM Higgs searches, 
from flavor and precision observables and from charge or color breaking (CCB) minima
on the phenomenological viability were highlighted. These results were obtained for a 
general super weakly interacting LSP\@.
Here, we will specify the LSP to be the gravitino which allows us to apply constraints from 
cosmological observations and conclude on the allowed values for the reheating temperature.
To this end we will extend the 17-dimensional parameter space introduced in 
\cite{Heisig:2013rya} by the additional parameter of the gravitino mass. Requiring that the LSP 
abundance matches the measured DM density we will compute the corresponding
reheating temperature by considering the thermal and non-thermal production of gravitinos. 
After computing the life-time and hadronic branching ratios of the stau
we will utilize the BBN bounds presented in \cite{Jedamzik:2007qk,Jedamzik:2006xz}.
We will choose the conservative values for ${\rm ^6Li/^7Li}$ here.
The analysis reveals the highest reheating temperatures that are consistent with 
bounds from BBN and other sensitive astrophysical observations, flavor and
precision bounds, theoretical bounds from vacuum stability, bounds from direct SUSY searches 
at the 7 and $8\TEV$ LHC as well as bounds from the MSSM Higgs searches and the 
requirement of providing a Higgs around $125\GEV$.
Our analysis shows that points with large $\TR$ as required by leptogenesis only survive in 
a very particular corner of the SUSY parameter space. Those spectra feature a distinct signature 
at colliders \cite{Lindert:2011td} that could be looked at in the upcoming LHC run. In particular, it 
requires the triggering on very slowly moving heavy stable charged particles (HSCP) which is 
expected to be challenging in the high-luminosity run. 

The paper is organized as follows. In section \ref{sec:gr_abundance} we will review 
the relevant production mechanisms of gravitinos and discuss the underlying assumptions 
made for the non-thermal production in our setup. In section \ref{sec:implLOSP} we will 
describe the cosmological implications of a late decaying stau that are relevant for our 
analysis. The computational steps of the pMSSM parameter scan are introduced in section 
\ref{sec:gr_genchain}. In section \ref{sec:gravres} we present our results and discuss the 
implications for the upcoming high-energy LHC run. We will conclude in section~\ref{sec:Conclusion}.

\section{Gravitino DM abundance} \label{sec:gr_abundance}

Recent measurement of the CMB power spectrum by the Planck satellite can be 
well described by the standard spatially flat \LCDM\ model with six cosmological 
parameters. Within this model the cold DM density has been measured with great 
precision \cite{Ade:2013zuv}. Combining the Planck power spectrum data with the 
WMAP polarization measurements \cite{Bennett:2012fp}, BAO measurements 
\cite{Percival:2009xn,Padmanabhan:2012hf,Blake:2011en,Anderson:2012sa,Beutler:2011hx}
as well as ground based high multipole measurements performed by the Atacama 
Cosmology Telescope \cite{Das:2013zf} and the South Pole Telescope \cite{Reichardt:2011yv} 
a best-fit value of
\beq
\label{eq:bestfitomegac}
\Omega_\text{CDM} h^2= 0.11889 
\eeq
was derived \cite{Ade:2013zuv}. This value will be considered for the 
following analysis.

There are two main production mechanisms for a gravitino which is not ultra-light 
and thus leads to a long-lived NLSP. On the one hand, this is the thermal production 
of gravitinos through inelastic scattering of particles participating in the thermal 
bath of the universe during the stage of reheating. On the other hand, it
is the non-thermal production through decays of metastable supersymmetric
particles into the gravitino.\footnote{Further sources of non-thermal production
could arise from the decay of the inflation field. Since this contribution depends
upon the actual model of inflation \cite{Giudice:1999am,Kallosh:1999jj}, 
we will not consider this contribution here.}

\subsection{Non-thermal production of gravitinos} \label{sec:nonthgrav}

\begin{figure}[thp]
\centering
\setlength{\unitlength}{1\textwidth}
\begin{picture}(0.42,0.36)
 \put(-0.024,0){ 
  \put(0.0,0.025){\includegraphics[scale=1.2]{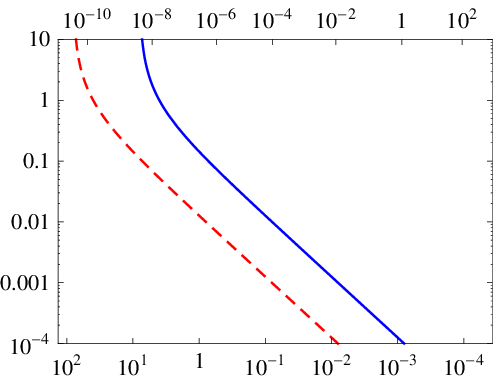}} 
   \put(0.095,0.08){\footnotesize $\mstau=2\TEV$}
   \put(0.23,0.17){\footnotesize $\mstau=200\GEV$}
  \put(0.18,0.0){\footnotesize $T_0\,[\GEV\,]$}
  \put(0.196,0.347){\footnotesize $t\,[\,\sec\,]$}
  \put(-0.038,0.076){\rotatebox{90}{\footnotesize $Y(T_0)/Y(T_0\to0)-1$}}
  }
\end{picture}
\caption{Stau yield as a function of the temperature $T_0=\mstau/x_0$ normalized
to its value at $T_0\to 0$ for the case of a typical annihilation process (see footnote
\ref{fn:typical}). We choose $\xf=25$ for this plot.
The upper axis labeling denotes the corresponding cosmic
time choosing $g_*(T)$ according to the particle content of the SM 
\cite{Kolb:1990vq}. By doing so we assume no additional relativistic degrees
of freedom for temperatures $T\lesssim10\GEV$ in our model.
}
\label{fig:Y-x0}
\end{figure}

In our setup the non-thermal production of gravitinos takes place via decays of
the stau into the gravitino. Due to the assumed $R$-parity conservation each 
stau eventually decays into a gravitino. Hence, the number density of staus 
before their decay, $n_{\stau_1}$, is equal to the number density of the gravitinos 
after all staus have decayed, $n_{\s G}$, and thus
\beq
\label{eq:nonthprod}
\Omega_{\s G}^{\text{non-th}} h^2 = \frac{\mG}{\mstau}\,\Omega_{\stau_1} h^2 \,.
\eeq
However, this picture only remains true, if the decay of the stau takes place separated 
from the efficient annihilation of the staus into SM particles, i.e., if these annihilation 
processes do not compete with the decay. In order to quantify this requirement we 
consider the stau yield, $Y=n_{\stau_1}/s$, where $s$ is the entropy density.
In figure \ref{fig:Y-x0} we show the evolution of the stau yield as a function of 
(decreasing) temperature $T_0$ and (increasing) time for a typical annihilation 
process\footnote{%
We choose an annihilation process for which the thermally averaged annihilation 
cross section, $\sveff$, can be expanded in $1/x\equiv T/\mstau$ as
\beq
\label{eq:expan}
\sveff= A \mstau^{-2} + \mathcal{O}\left(1/x\right)\,,
\eeq
where $A$ is dimensionless, containing only numerical factors, mixing angles, 
couplings and mass ratios, see e.g. \cite{Gondolo:1990dk,Griest:1990kh,Kolb:1990vq}. 
The first term in \eqref{eq:expan} often provides a good approximation
\cite{Gondolo:1990dk}.
The yield is then proportional to
\beq
Y(x_0)\propto\frac{\mstau}{\int_{\xf}^{x_0} \D x\, x^{-2} A} 
= \frac{\mstau}{A\left( x_{\text{f}}^{-1} - x_0^{-1}\right) }\,.
\label{eq:Yprop}
\eeq
For a fixed $\xf=\mstau/\Tf$, this expression uniquely determines the shape of the curves
in figure \ref{fig:Y-x0} independent of the considered process. Here, $\Tf$ is the freeze-out 
temperature which is typically of the order $\Tf\simeq\mstau/25$ \cite{Griest:1990kh}.
\label{fn:typical}
}
and for $\mstau=200\GEV$ and $2\TEV$. We plot the relative deviation of the yield from
its value for $T_0\to 0$ (if the stau were stable). This value is the quantity computed by 
\textsc{micrOMEGAs}~\cite{Belanger:2008sj} which will be used for our analysis.
For cosmic times after $10^{-4}\sec$ the deviation is around or below one percent.
Hence, for significantly smaller life-times of the stau, decays take 
place while significant annihilation processes are still ongoing. With respect to
the separated processes of annihilation and decay, this would lead to a higher
gravitino abundance and would require incorporating the stau decay term in the 
Boltzmann equations. However, in this work we will focus on stau lifetimes larger 
than $10^{-4}\sec$, first, because smaller life-times require gravitino masses
which are far too small to achieve high reheating temperatures as desired for
leptogenesis and thus are not of particular interest. Second, because BBN bounds 
that are subject to the investigation in this paper do not impose any restriction 
for lifetimes smaller than $10^{-2}\sec$. 

\subsection{Thermal production of gravitinos}

The relic abundance of thermally produced gravitinos, $\Omega_{\s G}^{\text{th}}$,
can be computed by solving the Boltzmann equation for the gravitino number density,
\beq
    \frac{\D n_{\s G}}{\D t} + 3 H n_{\s G} = C_{\s G}\,,
\label{eq:gravboltzmann}
\eeq
where the collision term $C_{\s G}$ is determined by the thermal gravitino production 
rates. It has been computed to leading order in the involved gauge couplings
considering the contribution from SUSY chromodynamics \cite{Bolz:2000fu} and the full 
SM gauge group \cite{Pradler:2006hh}. 
After the computation of $C_{\s G}$, \eqref{eq:gravboltzmann} can be solved analytically
and yields \cite{Pradler:2006qh}
\beq
\Omega_{\s G}^{\text{th}}h^2 =  \sum_{i=1}^{3} \omega_i\, g_i^2 
\left(1+\frac{M_i^2}{3\mG^2}\right) \log\left(\frac{k_i}{g_i}\right)
\left(\frac{\mG}{100\GEV}\right) \left(\frac{\TR}{10^{10}\GEV}\right),
\label{eq:omegagrthermal}
\eeq
where $g_i$ and $M_i$ are the gauge coupling and the gaugino mass
parameter, respectively, associated with the SM gauge groups $U(1)_Y$, 
$SU(2)_{\text{L}}$, $SU(3)_{\text{c}}$ and $k_i$, $\omega_i$ are corresponding 
numerical constants listed in table \ref{tab:const}. The couplings and gaugino 
mass parameters are understood to be evaluated at the scale $\TR$.
\begin{table}[h]
\begin{center}
\renewcommand{\arraystretch}{1.1}
\begin{tabular}{c | cccccc} 
gauge group & $i$ & $g_i$ & $M_i$  &   $k_i$ &  $\omega_i$ \\ \hline
$U(1)_Y$ & 1 & $g'$    & $M_1$  & 1.266  & 0.018 \\
$SU(2)_{\text{L}}$      & 2 & $g$     & $M_2$    & 1.312  & 0.044  \\
$SU(3)_{\text{c}}$ & 3 & $g_{\text{s}}$ & $M_3$    & 1.271  & 0.117
\end{tabular}
\end{center}
\caption{Assignments of the index $i$, the gauge coupling $g_i$, the gaugino mass 
parameter $M_i$ and the values of the associated constants $k_i$ and $\omega_i$ 
to the SM gauge groups $U(1)_Y$, $SU(2)_{\text{L}}$, and $SU(3)_{\text{c}}$.
Taken from \cite{Pradler:2006qh}.\label{tab:const}}
\end{table}

\section{Implications of the stau decay} \label{sec:implLOSP}

For a given MSSM parameter point all couplings of the gravitino to the MSSM particles 
are determined by the gravitino mass. We assume here that all heavier sparticles decay 
into the stau NLSP sufficiently fast so that direct decays of those sparticles into the gravitino 
are unimportant. The cosmological validity of a given parameter point then mainly depends 
on the yield, lifetime and the partial widths of the stau. 

For $\mstau-\mG>m_{\tau}$ the stau life-time, $\tau_{\stau_1}$, is dominated by the 
2-body decay $\stau_1\to\s G \tau$ which can be computed from the relevant terms 
in the interaction lagrangian of a massive spin-$3/2$ gravitino 
\cite{Bolz:2000fu,wess1992supersymmetry,Moroi:1995fs},
\beq
\mathcal{L}_{\text{int}}=-\frac{1}{\sqrt{2}\Mp}\left[ 
\overline{\tau} P_{\text{L}}\gamma^\mu \gamma^\nu\psi_\mu \left(\partial_\nu \stau_{\text{R}}\right) 
+ \overline{\tau} P_{\text{R}}\gamma^\mu \gamma^\nu\psi_\mu \left(\partial_\nu \stau_{\text{L}}\right)
\right]\,,
\eeq
where $\psi_\mu$ denotes the gravitino field and $\Mp$ is the reduced Planck mass. 
For the general case of non-vanishing left-right mixing in the stau sector, 
$\stau_1 = \cos\theta_\stau\,\stau_\text{L}+\sin\theta_\stau\,\stau_\text{R}$, we 
obtain the result
\beq
\begin{split}
 \tau_{\stau_1}^{-1}  \simeq \Gamma(\stau_1\to\s G \tau)
 = \;&\frac{\left( m_{\stau_1}^2 - \mG^2 - m_{\tau}^2 \right)^4 }{48\pi\Mp^2\mG^2m_{\stau_1}^3 }\,
\left[1+\frac{2\mG m_{\tau}\sin2\thest}{ m_{\stau_1}^2 - \mG^2 - m_{\tau}^2 } \right] \\
& \times\left[1-\left( \frac{2\mG m_{\tau}}{ m_{\stau_1}^2 - \mG^2 - m_{\tau}^2 }\right)^2 \right]^{3/2}.
\label{eq:taustau}
\end{split}
\eeq
The term proportional to $\sin2\thest$ (i.e., proportional to the amount of left-right mixing)
can become significant for small mass splittings between the stau and the gravitino. It 
leads to a decrease or increase of the life-time depending on the sign of $\sin2\thest$ 
which corresponds to the sign of  $-X_\tau = -A_\tau+\mu\tan\beta$ (see, e.g., appendix~B
in \cite{Heisig:2013rya}). This result reduces to the one given in \cite{Buchmuller:2004rq} 
for the case of a purely left- or right-handed stau, $\thest=0,\pi/2,\pi$, and is analogous to 
the result found in \cite{DiazCruz:2007fc} (published version) for the case of a stop NLSP\@.

The scenario is subject to several bounds. The most important bounds come from BBN 
constraints. The particles that are emitted in the decay of the stau into the gravitino can 
induce hadronic and electromagnetic showers at cosmic times characterized by the 
life-time of the stau. The produced energetic hadrons and photons induce hadro- and 
photodissociation processes that potentially distort the predictions for the light element 
abundances of standard BBN \cite{Jedamzik:2007qk,Jedamzik:2006xz,Kawasaki:2004yh,
Kawasaki:2004qu,Kawasaki:2008qe}. 
Furthermore, staus may form bound states with the background nuclei potentially leading 
to a catalyzed overproduction of ${}^6$Li \cite{Pospelov:2006sc,Pradler:2007is}. 
For the application of the BBN bounds it is crucial to determine the hadronic branching 
fractions. The tau emitted in the 2-body decay of the stau, $\stau_1\to\s G \tau$, has a 
hadronic branching fraction of roughly 65\%. However, for cosmic times up to about 
$3\SEC$ the interaction time of the tau is smaller than its life-time and the tau scatters
off the background before decaying. This scattering leads to a purely electromagnetic 
energy release \cite{Steffen:2006hw}. For later times the interaction time increases with 
decreasing temperature and hadronic decays of the tau become important. The mesons 
produced in the tau decays are unstable. 
In order to have a relevant effect on the BBN, the mesons have to scatter before their decay. 
This in turn only happens for cosmic times up to about $100\SEC$~\cite{Kawasaki:2004yh}.
For later times BBN constraints are dominated by nucleons emitted in the stau decay. 
These nucleons stem mainly from the 4-body decays $\stau_1\to\s G \tau q\bar q$ and 
$\stau_1\to \G \nu_{\tau} q\bar q^{\prime}$ with an 
invariant mass of the quark pair above the production threshold of the nucleon pair, 
$m_{q\bar q},m_{q\bar q^{\prime}}\gtrsim 2\GEV$~\cite{Steffen:2006hw}. 

If the life-time of the stau is very large, $\tau_{\stau_1}\gtrsim 10^{12}\SEC$, decays take 
place after the era of recombination and we can probe direct signatures of the stau decays 
in the measurements of the extra-galactic diffuse gamma ray background \cite{Kribs:1996ac}. 

For even larger life-times much stronger bounds can be obtained from the searches for 
anomalously heavy hydrogen in deep sea water \cite{Smith:1982qu,Hemmick:1989ns,
Yamagata:1993jq}. These measurements can be interpreted to provide a 95\% C.L. 
bound on the yield of charged relics today,
\beq
\label{eq:seawater1}
Y_{\text{today}} \lesssim 10^{-38}
\left(\frac{\Omega_{\text{B}}h^2}{0.022} 
\right)\,, 
\eeq
for the mass region $\mstau\le1600\GEV$ \cite{Yamagata:1993jq} and
\beq
\label{eq:seawater2}
Y_{\text{today}} \lesssim 10^{-32}
 \left(\frac{\Omega_{\text{B}}h^2}{0.022}
 \right)\,, 
\eeq
for the mass region $1600\GEV<\mstau\le2000\GEV$, where we chose an interpolated 
value between the ones given in \cite{Hemmick:1989ns} as an approximation.
The limits translate into a maximal life-time,
\beq
\label{eq:tauupperbound}
\tau_{\stau_1}<t_0 \left(\log \frac{Y}{Y^{\text{limit}}_{\text{today}}}\right)^{-1},
\eeq
where $t_0$ is the age of the universe, $t_0=4.354\times 10^{17}\sec$ \cite{Ade:2013ktc},
and $Y$ is the stau yield before their decay. We will only consider parameter points that 
obey \eqref{eq:tauupperbound} in the following analysis.\footnote{%
Note that the limit on the stau life-time  \eqref{eq:tauupperbound} depends only logarithmically 
on $Y^{\text{limit}}_{\text{today}}$. Moreover, we will use \eqref{eq:tauupperbound} only to 
determine an upper limit on the gravitino mass via \eqref{eq:taustau}. This upper limit is again 
not very sensitive to the exact value for the upper limit on $\tau_{\stau_1}$ (typically lying in the 
ballpark of $10^{16}\sec$) as large variations in the life-time correspond to very small 
variations in the mass gap between the stau and the gravitino in this region. Therefore our 
analysis is only sensitive to the rough order of magnitude of \eqref{eq:seawater1} and 
\eqref{eq:seawater2}.
}

Finally, we mention that one can also impose bounds on the life-time and abundance of
late decaying particles from the observation of the CMB. The secondary particles
produced in such a decay could affect the process of thermalization leading to 
a spectral distortion of the CMB away from a perfect black body spectrum 
\cite{Hu:1993gc,Feng:2003uy,Feng:2004mt,Lamon:2005jc,Chluba:2011hw}. 
However, the derivation and application of bounds from the CMB is beyond the 
scope of this work and is left for future investigations.

\section{Computational steps of the scan} \label{sec:gr_genchain}

\subsection{Scan over the 17-dimensional pMSSM}\label{sec:17pMSSM}

In this work we employ the Monte Carlo scan performed in \cite{Heisig:2013rya}. 
In this subsection we will briefly summarize the computational steps and the constraints 
imposed on the parameter space. For further details we refer to \cite{Heisig:2013rya}. 
We scanned over the 17-dimensional pMSSM parameter space with the following
input parameters and scan ranges:
\bea
-10^4\GEV \le &A_t &\le10^4\GEV \nonumber \\
-8000\GEV \le &A_b,\,A_\tau, \mu& \le8000\GEV \nonumber \\
1 \le & \tan\beta & \le 60 \nonumber \\
100\GEV \le &  m_A  & \le 4000\GEV \nonumber \\
200\GEV \le & \mstau & \le 2000\GEV \nonumber \\
\max(\mstau,700\GEV) \le & m_{\tilde{t}_1}, m_{\tilde{b}_1} & \le 5000\GEV \label{eq:scanranges}\\
0 < &  \thest, \theta_{\tilde{t}} & < \pi \nonumber \\
\mstau  \le & m_{\s L_{1,2}}, m_{\s e_{1,2}} & \le  4000\GEV \nonumber \\
\max(\mstau,1200\GEV)\le & \!\!m_{\s Q_{1,2}}\!\!
=  m_{\s u_{1,2}}\!= m_{\s d_{1,2}}\!\! & \le  8000\GEV \nonumber \\
\mstau  \le &  M_1, M_2 & \le  4000\GEV \nonumber \\
\max(\mstau,1000\GEV) \le & M_3 & \le  5000\GEV \nonumber 
\eea
For the third generation sfermions the spectrum parameters were chosen as input parameters. 
For simplicity we set $m_{\s Q_{1,2}}= m_{\s u_{1,2}}= m_{\s d_{1,2}}$. 
We imposed several hard constraints on the parameter space. The lighter stau was taken to 
be the NLSP\@, hence we only accepted points where
\beq
\label{eq:stauLOSPcon}
\stau_1 = \text{NLSP}\,.
\eeq
Further, we required that at least one of the neutral $CP$-even Higgses, $m_h,m_H$, can 
be identified with the recently discovered Higgs boson at the LHC
\cite{ATLAS-CONF-2013-014,CMS-PAS-HIG-13-005},
\beq
\label{eq:higgswindow}
m_h\;\,\text{or/and} \;\,m_H \in [123;128]\GEV.
\eeq

We generated the sparticle spectrum with \textsc{SuSpect}~2.41 \cite{Djouadi:2002ze}. 
For the third generation sfermions we used tree-level relations in order to translate the 
chosen input parameters into soft parameters that feed into the spectrum generator.
The Higgs sector was recalculated using \textsc{FeynHiggs}~2.9.2~\cite{Heinemeyer:1998yj}.
We computed the stau yield with \textsc{micrOMEGAs}~2.4.5~\cite{Belanger:2008sj}. 

We imposed several experimental and theoretical constraints on the parameter space. 
Lower bounds on the sparticle masses were derived from searches for heavy stable 
charged particles (HSCP) at the LHC\@. To this end and in order to discuss the
perspective for a future discovery at the LHC, we determined all relevant cross sections 
for a center-of-mass energy of 7, 8 and $14\TEV$. We computed the direct stau production 
via $s$-channel Higgses $h,H$ with \textsc{Whizard} 2.1.1 \cite{Kilian:2007gr}.
The cross sections for all other contributions were estimated via a fast interpolation
method using grids computed with \textsc{Prospino}~2.1~\cite{1997NuPhB.492...51B,
1999PhRvL..83.3780B,Plehn:2004rp,Beenakker:1997ut}
as well as grids from the program package \textsc{NLLfast} \cite{Beenakker:2009ha,
Beenakker:2010nq,Kulesza:2008jb,Kulesza:2009kq}. The cross section upper limits 
were estimated from a reinterpretation of the HSCP searches for the 7 and $8\TEV$ runs 
reported by CMS \cite{CMS1305.0491}. For spectra with mass-degenerate staus and 
colored sparticles the respective $R$-hadron searches were taken into account.
The decay widths and branching ratios were computed with 
\textsc{SDecay}~\cite{Djouadi:2006bz,Kraml:2007sx}
and \textsc{Whizard}~2.1.1~\cite{Kilian:2007gr}.

We considered bounds from flavor and precision observables. We applied the constraints 
$\text{BR}(B\to X_s\gamma) \in [2.87 ; 3.99] \times 10^{-4}$ \cite{HFAGbsgAug12} and 
$\text{BR}(B_s^0\to\mu^+\mu^-) \in [1.1 ; 6.4] \times 10^{-9}$  \cite{:2012ct} on the respective 
observables computed by \textsc{micrOMEGAs}~2.4.5~\cite{Belanger:2008sj}. Constraints 
on the corrections to the mass of the $W$ boson were taken into account by applying the limit 
$M_W \in [80.325 ; 80.445] \GEV$ \cite{Group:2012gb,Bechtle:2012jw,Heinemeyer:2006px} 
to the value calculated by \FH~2.9.2. For the computation of exclusion bounds from collider 
searches for the MSSM Higgs sector, performed at LEP, the Tevatron and the LHC, we utilized 
\textsc{HiggsBounds 4.0.0}~\cite{Bechtle:2011sb}.
Theoretical constrains from charge or color breaking (CCB) minima were taken into account by
applying upper bounds on $|\mu\tan\beta|$ \cite{Kitahara:2013lfa} and $|A_\tau|,|A_b|,|A_t|$ 
\cite{Frere:1983ag,AlvarezGaume:1983gj,Claudson:1983et,Kounnas:1983td,Derendinger:1983bz}.

The point density was adjusted to the expected variation of the yield. In co-anni\-hilation regions
and regions around resonances or thresholds proportionally more points were accumulated 
(see \cite{Heisig:2013rya} for details). We use a set of $10^6$ pMSSM scan points\footnote{%
With additional computing time the number of scan points was doubled
with respect to \cite{Heisig:2013rya}. However, the composition of points remains
unchanged.
} 
obeying the hard constraints \eqref{eq:stauLOSPcon} and \eqref{eq:higgswindow}.

\subsection{Extension of the pMSSM parameter scan}

We will now extend the 17-dimensional pMSSM scan described in \cite{Heisig:2013rya}
incorporating the gravitino LSP\@. For each point of the 17-dimensional pMSSM 
parameter space we perform the following computational steps. First, we determine the 
possible mass range for the gravitino under the following restrictions depending on the
stau mass, the stau mixing angle and the yield of the given parameter point. The resulting 
life-time of the stau is required to be greater than $10^{-4}\SEC$---motivated by the 
arguments given in section \ref{sec:nonthgrav}---and smaller than the upper bound from 
\eqref{eq:tauupperbound}. From \eqref{eq:taustau} this imposes a lower and upper bound 
on the gravitino mass. Furthermore, the non-thermal contribution to the gravitino abundance 
\eqref{eq:nonthprod} should not exceed the measured DM abundance (see below for further 
details). This requirement imposes an additional upper limit on the gravitino mass which can 
be both either more or less restrictive than the upper bound from \eqref{eq:tauupperbound}. 
Second, for a given point we randomly generate 10 values for $\mG$ in the required 
interval. Since the interval spans over several orders of magnitude we use logarithmic
priors here. The following steps are then performed for each of the 10 gravitino mass
points. 

We computed the non-thermal contribution to the gravitino abundance from the
stau yield with \eqref{eq:nonthprod}. By demanding that the resulting total gravitino 
abundance matches the measured DM abundance, 
$\Omega_{\s G}^{\text{non-th}} h^2+\Omega^{\text{th}}_{\s G}h^2
=\Omega_\text{CDM}h^2$, we compute the required
abundance of thermally produced gravitinos\footnote{%
Note, that the result \eqref{eq:omegagrthermal} was obtained using hard thermal loop 
resummation \cite{Braaten:1989mz} which requires weak couplings. Hence, the result 
might not be reliable for small reheating temperatures 
$\TR\lesssim 10^6\GEV$~\cite{Pradler:2006qh}.},
$\Omega^{\text{th}}_{\s G}h^2$. For $\Omega_\text{CDM}h^2$, we chose the best-fit 
value \eqref{eq:bestfitomegac}.\footnote{%
The $68\%$ confidence interval for the $\Omega_\text{CDM}h^2$ \cite{Ade:2013zuv} is 
much smaller than the expected precision of the computations performed
here. Therefore, we refrain from varying the $\Omega_\text{CDM}h^2$
within the confidence interval by a Monte Carlo method. The effect of
such a treatment would be marginal.} 
From \eqref{eq:omegagrthermal} we compute the reheating temperature, $\TR$,
that provides $\Omega^{\text{th}}_{\s G}h^2$ for the given parameter point.
Since $M_i$ and $g_i$ have to be evaluated at the scale $\TR$, these quantities 
are functions of $\TR$ and the equation has to be solved iteratively. However, we 
achieved a fast convergence within 2 to 4 iterations to a more than sufficient accuracy.
For the evaluation of $g_i$ and $M_i$ we take into account the one-loop running
\beq
\label{eq:gaugerun}
g_i(\TR) 
= \frac{g_i(Q_{\text{in}})}{\sqrt{1-\frac{b_ig_i^2(Q_{\text{in}})}{8\pi^2} \log\left(\frac{\TR}{Q_{\text{in}}}\right)}}\,,
\eeq
and the fact that 
\beq
\label{eq:nonren}
M_i(\TR) = \left(\frac{g_i(\TR)}{g_i(Q_{\text{in}})}\right)^2M_i(Q_{\text{in}})\,,
\eeq
see e.g.~\cite{drees2004theory}. In \eqref{eq:gaugerun}, $b_i$ are the 
MSSM coefficients of the 1-loop renormalization group equations, 
$(b_1,b_2,b_3) = (11,1,-3)$ and $Q_{\text{in}}$ is the input scale, which we 
choose to be the electroweak scale here.\footnote{%
We tolerate a slight overestimation of the couplings $g_i(\TR)$ that could arise 
from the fact that the running with the MSSM coefficients starts below the precise 
mass scale of the corresponding SUSY particles. The effect on the final results 
is, however,  expected to be marginal.}

For the interpretation of BBN bounds and bounds from diffuse gamma ray
observations we compute the life-time, \eqref{eq:taustau}, and the hadronic 
branching ratios, $B_{\text{h}}$, of the stau. For $\tau_{\stau_1}\gtrsim100\SEC$ the
relevant contributions to $B_{\text{h}}$ stem from 4-body decays,
\beq
B_{\text{h}}=\frac{\Gamma(\stau_1\to \s G \tau q\bar q)
+\Gamma(\stau_1\to \G \nu_{\tau} q\bar q^{\prime})}{\Gamma_{\text{tot}}}\,,
\eeq
where $\Gamma_{\text{tot}}$ is the total width, which we approximate by the 
2-body decay, $\Gamma(\stau_1\to \s G \tau)$ being the dominant decay mode.
The partial widths $\Gamma(\stau_1\to \s G \tau q\bar q)$ and 
$\Gamma(\stau_1\to \G \nu_{\tau} q\bar q^{\prime})$ include the decays into 
all kinematically accessible quark-anti-quark pairs. However, the contributions from 
diagrams containing top quarks in the final state are found to be negligible for all 
situations relevant here. We perform the computation of $B_{\text{h}}$ with the
spin-$3/2$ extension of \textsc{HELAS} \cite{Hagiwara:2010pi} implemented 
in \textsc{MadGraph} \cite{alwall-2007}.
This program package supports the computation of arbitrary tree-level amplitudes 
with external gravitinos interacting with MSSM particles. In order to save computing 
time we determine the hadronic branching ratios in two steps on an increasing level 
of accuracy.

In the first step we conservatively estimate $B_{\text{h}}$ on the basis of a precomputed 
grid. To this end we computed $B_{\text{h}}$ as a function of the stau life-time for various 
choices of the stau masses and use an interpolation routine to obtain the values for 
arbitrary masses. For the computation of the grid we ignored left-right mixing effects and 
considered a purely right-handed stau taking into account diagrams with 
$Z/\gamma$-exchange only. Equally, we set the masses of all sparticles heavier than the 
stau to $3\mstau$. This way diagrams involving EWinos (and squarks) are suppressed 
and do not contribute. Those diagrams can potentially increase the hadronic branching 
ratios. As an example, in the case of a right-handed stau with $\mstau=500\GEV$ and 
$\mG=100\GEV$ we found a maximal enhancement of $B_{\text{h}}$ for almost 
mass-degenerate squarks of the first two generations and the bino-like neutralino, 
$\msq\simeq\mne\simeq510\GEV$, by a factor of three. 
The branching ratios computed in this way are in rough agreement with results found 
earlier \cite{Steffen:2006hw,Feng:2004zu}.\footnote{%
In \cite{Feng:2004zu} smaller hadronic branching ratios are achieved. This is expected
to stem from the photon interference which is not included in that computation as pointed 
out in \cite{Steffen:2006hw}. Our results are similar to those given in \cite{Steffen:2006hw} 
which are obtained, however, for $m_{\s B}=1.1\mstau$. Since in \cite{Steffen:2006hw} the 
results are shown as iso-$\mG$ curves in the $\mstau$-$B_{\text{h}}$ plane it is difficult 
to resolve the exact behavior of $B_{\text{h}}$ in the region of large life-times from the 
plot given in this reference.
}

In the second step, for each point that passes the bounds described in 
section~\ref{sec:17pMSSM} as well as the BBN bounds described further below 
(under the assumption of the conservatively estimated $B_{\text{h}}$) we 
recompute the hadronic branching ratios with \textsc{MadGraph} from the full spectrum.
To this end we consider all diagrams of the processes $\stau_1\to \G \tau  q\bar q$ and 
$\stau_1\to \G \nu_{\tau} q\bar q^{\prime}$ containing an intermediate vector boson, 
an intermediate light or heavy Higgs (for the process $\stau_1\to \G \tau b\bar b$) 
as well as all diagrams containing an intermediate lightest neutralino or chargino.
For a large fraction of scan points the contribution from 
$\stau_1\to \G \nu_{\tau} q\bar q^{\prime}$---mediated via $W^{\pm}$- or 
$\s\chi^{\pm}$-exchange---is found to be the most important. It can exceed the contribution 
from $\stau_1\to \G \tau  q\bar q$ ($q=d,u,s,c$) by up to an order of magnitude. The 
contribution from $\stau_1\to \G \tau b\bar b$ is less important in our scan and we found 
$\Gamma(\stau_1\to \G \tau b\bar b)/\Gamma(\stau_1\to \G \tau q\bar q)\simeq3$ at most,
where $q=d,u,s,c$ again. This contribution can potentially be enhanced from a Higgs 
exchange in the presence of large stau-Higgs couplings.
As argued above for all computations we impose the lower cut on the invariant mass 
of the quark pairs $m_{q\bar q},m_{q\bar q^{\prime}}>2\GEV$.

For life-times $\tau_{\stau_1}\lesssim100\SEC$ the interactions of the
mesons produced in the decays of the tau can become important. We estimate the
corresponding hadronic branching ratio by using the results given in \cite{Feng:2004zu}.

We apply the constraints from BBN derived in \cite{Jedamzik:2007qk,Jedamzik:2006xz}. 
This analysis takes into account effects from proton-neutron interconversion, hadro- and
photodissociation as well as all currently known bound-states effects. The constraints are
based on the following observationally determined limits on the light element abundances:
\bea
& Y_{p} < 0.258 \nonumber \\
& 1.2 \times 10^{-5} < {\rm ^2H/H} < 5.3 \times 10^{-5} \nonumber \\
& {\rm ^3He/^2H} <1.52 \label{eq:premabobs} \\
& 8.5 \times 10^{-11} < {\rm ^7Li/H} < 5 \times 10^{-10}\nonumber  \\
& \;{\rm ^6Li/^7Li} < 0.66\,.\nonumber
\eea
Here a conservative choice was made concerning the value of ${\rm ^6Li/^7Li}$.
As the BBN bounds derived in these references are  given in terms of the life-time of
the relic, its mass and its hadronic branching ratio, we do not compute the
hadronic energy release nor simulate the hadronization of primary partons here.
Rather we directly apply the computed values for $\tau_{\stau}$, $B_{\text{h}}$ to the
bounds given in \cite{Jedamzik:2007qk,Jedamzik:2006xz}. 
These bounds are given for two masses of the relic $m_X = 100\GEV,1\TEV$ and 
for (at least) six values for $B_{\text{h}}$ as a function of the life-time of the relic $\tau_X$. 
For life-times below $10^{7}\SEC$, where the hadronic energy release is 
important, the maximal yield which is compatible with the bounds, $Y_{\max}$,
almost scales like $B_{\text{h}}^{-1}$ and $m_X^{-1}$. Therefore we apply a linear 
interpolation (and extrapolation for masses above $1\TEV$) in $\log(B_{\text{h}})$ 
and $\log(m_X)$ between the corresponding values of $Y_{\max}$ for a given life-time.
We take the bounds for $10^2\SEC<\tau_X<10^9\SEC$ from \cite{Jedamzik:2007qk} 
(erratum from 2009). As the bounds in \cite{Jedamzik:2007qk} are only given for this interval,
for life-times $10^{-2}\SEC<\tau_X<10^2\SEC$ and $10^{9}\SEC<\tau_X<10^{12}\SEC$ 
we estimate the constraints by using the results of \cite{Jedamzik:2006xz}, 
where we ignored the curves for $B_{\text{h}}>0.01$ in the latter interval. 
The constraints in this analysis are, however, derived for a neutral relic. As stated in
\cite{Jedamzik:2007qk}, for large $B_{\text{h}}$---typically achieved for very small 
life-times---the constraints on charged and neutral particles are almost identical.
This is why we expect the analysis to apply for the former interval. For life-times in
the latter interval, effects of photodissociation are the most relevant effects 
from the decaying staus. We expect the corresponding constraint to be similar to the 
bounds on decaying neutral relics for $B_{\text{h}}>0.01$, which is indeed the case 
for life-times $10^8\SEC<\tau_X<10^9\SEC$ for which the constraints are given in 
both analyses.

For very large life-times $\tau_{\stau_1}>10^{12}\SEC$ we consider bounds derived from 
the observation of diffuse gamma ray emissions \cite{Sreekumar:1997yg}. We apply the 
relic density bounds for 2-body radiative decays derived in \cite{Kribs:1996ac}. These 
bounds become restrictive only for life-times of $\tau_{\stau_1}\gtrsim5\times10^{12}\SEC$ 
which corresponds to a mass splitting $\mstau-\mG\lesssim10\GEV$ in the considered scan 
region for $\mstau$. Consequently, the electromagnetic energy release in the stau decay 
is relatively small. We estimate the electromagnetic injection energy times photon branching 
ratio by
\beq
E_{\text{inj}} B_{\gamma} = 0.3\,\frac{\mstau^2-\mG^2}{2\mstau}\,,
\eeq
where the pre-factor 0.3 conservatively takes into account the energy 
taken away by neutrinos emitted in the tau decays \cite{Ellis:2003dn}.
In the most relevant region $10^{13}\SEC\lesssim\tau_{\stau_1}\lesssim10^{15}\SEC$
the constraints on $Y E_{\text{inj}} B_{\gamma}$ grow almost linear in $E_{\text{inj}}$ 
for small $E_{\text{inj}}$, i.e., the displayed curves for $E_{\text{inj}}=25\GEV$,
$50\GEV$ and $100\GEV$ are almost identical for these life-times. Assuming 
a linearity down to even smaller $E_{\text{inj}}$, we apply the limits
for the smallest value for the injection energy given, $E_{\text{inj}}=25\GEV$.\footnote{%
A similar analysis was applied in \cite{CahillRowley:2012cb}.}

\section{Results and discussion} \label{sec:gravres}

\begin{figure}[h!]
\centering
\setlength{\unitlength}{1\textwidth}
\begin{picture}(0.925,0.4)
 \put(-0.024,0){ 
  \put(-0.03,0.025){\includegraphics[scale=1.2]{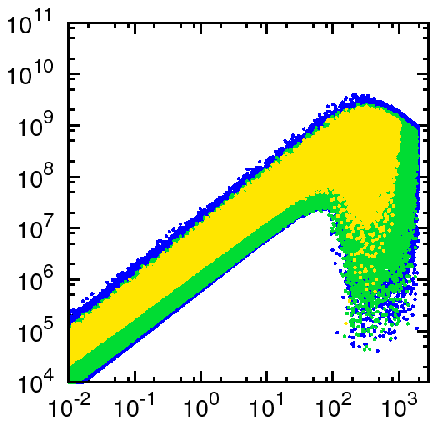} } 
  \put(0.19,0.0){\footnotesize $\mG\,[\GEV\,]$}
  \put(0.0,0.165){\rotatebox{90}{\footnotesize $\TR\,[\GEV\,]$}}
  }
 \put(0.445,0){  
  \put(0.0,0.025){\includegraphics[scale=1.2]{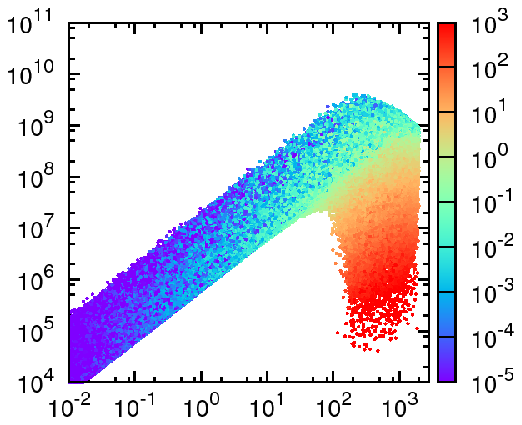}} 
  \put(0.19,0.0){\footnotesize $\mG\,[\GEV\,]$}
  \put(0.0,0.165){\rotatebox{90}{\footnotesize $\TR\,[\GEV\,]$}}
  \put(0.475,0.273){\rotatebox{-90}{\footnotesize $\Omega_{\s G}^{\text{non-th}}/\Omega_{\s G}^{\text{th}}$}}
  }
\end{picture}
\caption{Points of the $(17+1)$-dimensional pMSSM scan in the $\mG$-$\TR$ plane. 
Left panel:~Dominant contribution to the thermal gravitino production associated with $M_3$ 
(blue points), $M_2$ (green points) and $M_1$ (yellow points).
Right panel:~Ratio between the non-thermal and thermal contribution to the 
gravitino abundance, $\Omega_{\s G}^{\text{non-th}}/\Omega_{\s G}^{\text{th}}$.
}
\label{fig:grav3}
\end{figure}

The left panel of figure \ref{fig:grav3} shows the domains of the contributions to thermal 
gravitino production associated 
with the different gauge couplings. In blue, green and yellow we plotted points where the 
$SU(3)_{\text{c}}$, $SU(2)_{\text{L}}$ and $U(1)_Y$ contributions are dominant, respectively.
Note that the point density is saturated in large regions of the plane such that blue 
points are covered by green points etc. It is interesting to observe that all three contributions 
are important in our scan despite the smaller gauge coupling and numerical constant
$\omega_i$ for the $U(1)_Y$ contribution. However, the term associated with $U(1)_Y$ only 
provides a dominant contribution in a narrow band. This can be understood as follows.
For a given gravitino mass, points with a larger $\TR$ tend to have a lighter spectrum. 
The uppermost points are those where $M_1$, $M_2$ and $M_3$ are all close to the lower 
end of their scan interval. Since the scan range for the gluino mass parameter, $M_3$, starts 
at a larger value (in accordance with stronger mass bounds expected) for the uppermost stripe 
of the band the contributions from $M_1$ and $M_2$ are less important. On the other hand, 
scan points at the lowermost part of the band are those where $M_1$, $M_2$ and $M_3$ are 
all maximal. Moreover, we allow for slightly larger values for $M_3$ than for $M_1$ and 
$M_2$ in our scan. As a consequence the running of $M_1$---potentially rendering 
$M_1(\TR)\!>\!M_3(\TR)$---cannot compensate the smaller coupling and thus the 
$SU(3)_{\text{c}}$ and $SU(2)_{\text{L}}$ contributions are again the most important ones.

The right panel of figure \ref{fig:grav3} shows the ratio between the non-thermal and the 
thermal production of gravitinos. For small $\mG$ the non-thermal contribution is unimportant 
and the band spanned by the resulting reheating temperature grows linearly with the gravitino 
mass. Once the gravitino mass approaches the mass of the other superpartners we encounter 
two effects. First, according to \eqref{eq:omegagrthermal}, the linear growth of $\TR$ turns into 
a decrease when approaching small mass splittings between the gravitino and the gaugino 
masses. This effect causes the points with the highest $\TR$ to lie around gravitino masses 
of a few hundred GeV. The maximal $\TR$ reached by the generated points in our scan 
depends on the lower limits of the scan ranges for the gaugino masses, in particular for 
$M_3$.\footnote{%
Upper bounds on the gluino mass from the over-closure constraint were discussed in 
\cite{Fujii:2003nr,Steffen:2008bt,Endo:2010ya,Endo:2011uw}, see also 
\cite{Pradler:2006hh,Pradler:2006qh} for a discussion in the framework of constraint models.
}
Here, having chosen $M_3>1\TEV$, it reaches $\TR\simeq4\times10^9\GEV$ in accordance 
with the conservative limits found in \cite{Endo:2011uw}. As a second effect, once the 
gravitino approaches the stau mass non-thermal contributions become important. Depending 
on the stau yield of a considered point the required reheating temperature is pushed down 
by a more or less significant amount. The points that still lie within the linearly rising band 
when $\mG$ approaches $\mstau$ tend to be those with rather small yields. However,  we 
found points with yields $Y\gtrsim10^{-13}$ for $\TR\gtrsim10^9\GEV$. For these points the 
non-thermal contribution to the gravitino production is of the same order of magnitude as the 
thermal contribution and cannot be neglected. 

\begin{figure}[t]
\centering
\setlength{\unitlength}{1\textwidth}
\begin{picture}(0.9,0.4)
 \put(-0.024,0){ 
  \put(-0.03,0.025){\includegraphics[scale=1.21]{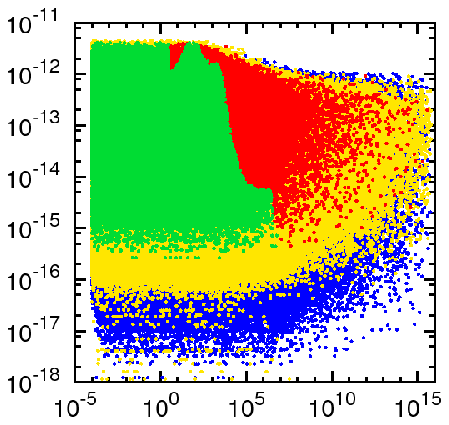} } 
  \put(0.205,0.0){\footnotesize $\tau_{\stau_1}\,[\,\sec\,]$}
  \put(0.0,0.2){\rotatebox{90}{\footnotesize $Y$}}
  }
 \put(0.46,0){  
  \put(-0.03,0.025){\includegraphics[scale=1.21]{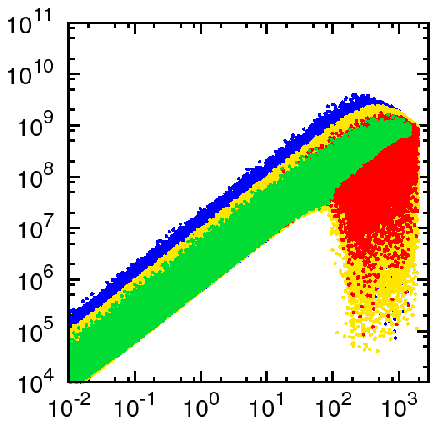}} 
  \put(0.19,0.0){\footnotesize $\mG\,[\GEV\,]$}
  \put(0.0,0.165){\rotatebox{90}{\footnotesize $\TR\,[\GEV\,]$}}
  }
\end{picture}
\caption{Points of the $(17+1)$-dimensional pMSSM scan. The color code is chosen as follows. 
Blue:~Points passing no constraints. Yellow:~Points passing constraints from the 
HSCP search. Red:~Points additionally passing the constraints from flavor and precision 
observables, \textsc{HiggsBounds} and CCB bounds. Green:~Points additionally passing 
the BBN bounds and bounds from the diffuse gamma ray spectrum. Left panel:~The stau yield 
$Y_{\stau_1}$ against the stau life-time $\tau_{\stau_1}$. Right panel:~Reheating temperature 
$\TR$ against the gravitino mass $\mG$. Note that the formation of horizontal lines in the left 
panel is a remnant of the scan, generating ten gravitino masses per point in the 17-dimensional 
pMSSM scan, all having the same $Y_{\stau_1}$ but different $\tau_{\stau_1}$.
}
\label{fig:grav1}
\end{figure}

\begin{figure}[h!]
\centering
\setlength{\unitlength}{1\textwidth}
\begin{picture}(0.9,0.4)
 \put(-0.024,0){ 
  \put(-0.03,0.025){\includegraphics[scale=1.2]{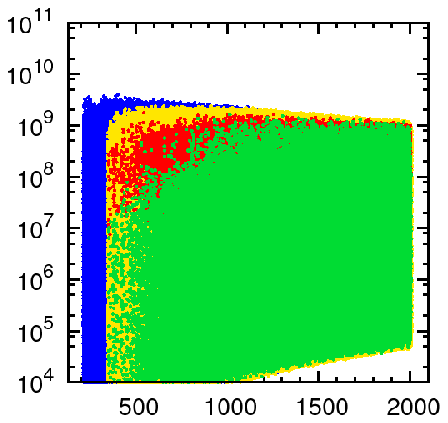}} 
  \put(0.19,0.0){\footnotesize $\mstau\,[\GEV\,]$}
  \put(0.0,0.165){\rotatebox{90}{\footnotesize $\TR\,[\GEV\,]$}}
  }
 \put(0.46,0){  
  \put(-0.03,0.025){\includegraphics[scale=1.2]{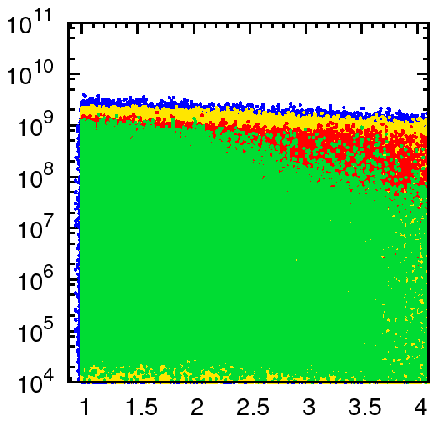} } 
  \put(0.205,0.0){\footnotesize $M_2/\mstau$}
  \put(0.0,0.165){\rotatebox{90}{\footnotesize $\TR\,[\GEV\,]$}}
  }
\end{picture}
\caption{Points of the $(17+1)$-dimensional pMSSM scan. Color code as in figure 
\ref{fig:grav1}. Left panel:~Reheating temperature $\TR$ against the stau
mass $\mstau$. Right panel:~Reheating temperature $\TR$ against the mass ratio
$M_2/\mstau$. 
}
\label{fig:grav2}
\end{figure}

In figure \ref{fig:grav1} we show the effect of the bounds imposed on the 
$(17+1)$-dimensional parameter space in the $\tau_{\stau_1}$-$Y_{\stau_1}$ 
plane and in the $\mG$-$\TR$ plane. The blue and yellow points are rejected by the HSCP 
searches and by the additional bounds from flavor and precision observables, \HB\ and 
CCB bounds, respectively, as they have been described in section \ref{sec:17pMSSM}.
The red points are rejected by the BBN bounds or the bounds from the diffuse gamma ray 
spectrum.
The left panel of figure \ref{fig:grav1} reveals the effect of the BBN bounds on our 
parameter space. The border-line between the green and red points falls down relatively 
rapidly for life-times above $1000\SEC$ according to the stronger bounds from 
hadrodissociation processes as well as bound-state effects. For life-times above $10^6\SEC$ 
photodissociation processes become most restrictive. As a consequence we do not find 
allowed points with $\tau_{\stau_1}>10^7\SEC$ in our scan. However, the point density starts 
to dilute for $\tau_{\stau_1}>10^7\SEC$ as a consequence of our logarithmic prior in the scan 
over the gravitino mass (rather than over the stau life-time). Further, we do not encounter any 
point which is allowed by all other constraints but lies close to the bound on the yield imposed
by the diffuse gamma ray spectrum.
The spot of red points in the region $Y\gtrsim10^{-12}$ and $\tau_{\stau_1}\lesssim10^2\SEC$ 
stems from the energy release of mesons originating from tau decays.

Note that the BBN constraints from \cite{Jedamzik:2007qk} show almost no dependents on 
the hadronic branching ratios for $\tau_{\stau_1}>10^5\sec$ and for the typically achieved 
hadronic branching ratios in this region that are well below $B_h=10^{-2}$. Hence, the 
BBN constraints are not sensitive to the precise computation of $B_h$ in this region.

The right panel of figure~\ref{fig:grav1} shows the parameter points in the $\mG$-$\TR$ plane. 
The search for HSCP at the 7 and $8\TEV$ LHC imposes very restrictive limits on the gluino and wino 
masses, e.g., conservatively $\mgo\gtrsim1.2\TEV, \,M_2\gtrsim800\GEV$
\cite{Heisig:2013rya}.\footnote{%
The given numbers are conservative lower bounds (at 95\% C.L.). The bounds become considerably 
stronger if the production of several sparticles contribute equally strong to the production at the LHC\@. 
In our analysis all relevant channels are taken into account.
}
These limits exclude all points with a reheating temperature above $\TR\simeq2.3\times10^9\GEV$ 
(cf. blue versus yellow points). Bounds from flavor and precision observables, MSSM Higgs searches 
and CCB vacua further reduce the parameter space leaving a maximal reheating temperature of 
slightly below $2\times10^9\GEV$ (cf. yellow versus red points). The application of BBN bounds has 
the most significant effect in the region of large $\Omega_{\s G}^{\text{non-th}}/\Omega_{\s G}^{\text{th}}$ 
where $Y$ and $\mG$ are large.

The analysis reveals the existence of points which provide reheating temperatures
$\TR>10^9\GEV$ and are consistent with all discussed bounds and with a Higgs mass 
of around $125\GEV$. 
All these points share very distinct features. First, these points feature a heavy 
gravitino, $300\GEV<\mG<1.4\TEV$, resulting in a relatively large stau life-time,
$10^4\SEC<\tau_{\stau_1}<10^7\SEC$. 
It is interesting to note that the upper bound on the life-time (coming from BBN bounds) 
still causes a separation of the stau and gravitino masses of at least $200\GEV$ in our scan. 
Second, all points lie within the resonance region where $ m_A\simeq2\mstau$. In this region
exceptionally small stau yields can be achieved due to annihilation via a resonant $s$-channel 
heavy Higgs. For most points (88 points) the dominant annihilation process is resonant 
stau-pair annihilation \cite{Pradler:2008qc}.\footnote{%
The potential for large reheating temperatures $\TR>10^9\GEV$ in the region of resonant
stau-pair annihilation was also found in \cite{Endo:2011uw}. In this reference conservative upper limits
on the gluino mass were obtained for two different choices of the bino and wino mass parameters
and in three different regions of dominant stau annihilation processes: dominant electroweak
annihilation of staus, annihilation into light Higgses \cite{Pradler:2008qc,Ratz:2008qh} 
and stau annihilation via a resonant heavy Higgs \cite{Pradler:2008qc}. After applying BBN and 
CCB constraints the loosest limits on the gluino mass were obtained in the latter region. 
}
For three points effects of co-annihilation are important: we found that one and two points feature 
resonant stop and EWino co-annihilation \cite{Heisig:2013rya} as the dominant annihilation process, 
respectively. Note that EWino co-annihilation via a resonant heavy Higgs requires no particularly large 
Higgs-sfermion couplings. Thus, the viability of these points does not depend upon constraints from 
CCB vacua. 

Third, for most points the yield is smaller than $10^{-14}$. However, we encountered a few points 
with $10^{-14}<Y<3\times10^{-14}$. In order to compensate for the slightly larger contribution of 
non-thermal gravitino production, those points were driven into a region of small gaugino masses
and thus very small mass splittings between the stau and the gauginos, $M_2/\mstau<1.2$, 
$M_1/\mstau<1.3$ and $M_3/\mstau<1.5$. 
This strong tendency for small gaugino masses is in fact relaxed for $Y<10^{-14}$.  Still, we found 
no points with $M_2>2.1\mstau$ (cf. right panel of figure~\ref{fig:grav2}), $M_1>3.1\mstau$ or 
$M_3>3.7\mstau$ for $\TR>10^9\GEV$. The fact that (at the low scale) $M_1$ and $M_3$ are 
less constrained than $M_2$ is due to the smaller coupling in the former case and due to the 
slower running up to the scale $\TR$ in the latter case. 
The tendency for small stau-gaugino mass splittings is in fact the result of two effects.
On the one hand, according to \eqref{eq:omegagrthermal}, the gravitino mass that maximizes the 
reheating temperature for a given $\Omega_{\s G}^{\text{th}}$ grows with increasing gaugino 
masses. On the other hand, the preference for smaller stau life-times from BBN bounds favors
larger mass splittings between the stau and the gravitino. 
As a consequence the strong bounds on $\mgo$ and $M_2$ also lift up the stau masses for points 
with $\TR>10^9\GEV$ in our scan, which we found to lie above $\mstau\simeq800\GEV$ (see left 
panel of figure~\ref{fig:grav2}).

\begin{figure}[h!]
\centering
\setlength{\unitlength}{1\textwidth}
\begin{picture}(0.95,0.4)
 \put(-0.024,0){ 
  \put(0.0,0.025){\includegraphics[scale=1.18]{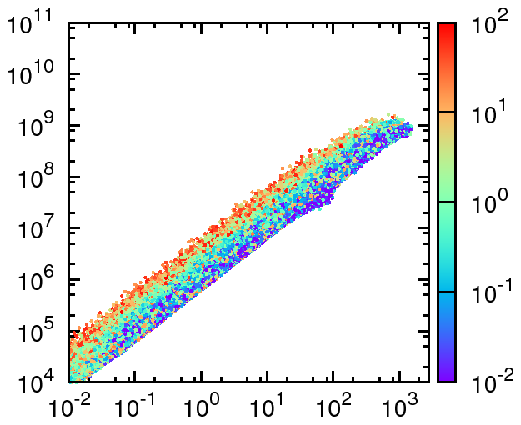}} 
  \put(0.18,0.0){\footnotesize $\mG\,[\GEV\,]$}
  \put(0.004,0.165){\rotatebox{90}{\footnotesize $\TR\,[\GEV\,]$}}
  \put(0.46,0.265){\rotatebox{-90}{\footnotesize $\sigma_{14\TEV}\;[\fb\,]$}}
  }
 \put(0.492,0){ 
  \put(0.0,0.025){\includegraphics[scale=1.18]{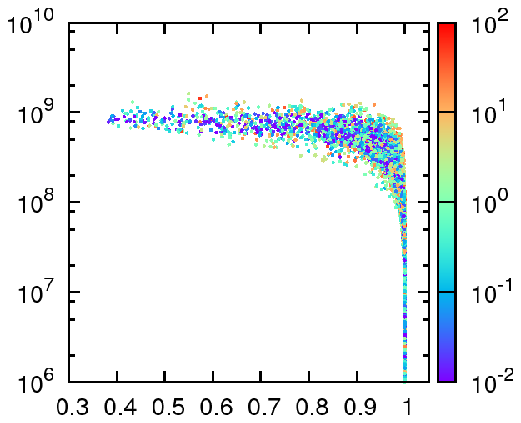}} 
  \put(0.17,0.0){\footnotesize $1-\mG^2/\mstau^2$}
  \put(0.004,0.165){\rotatebox{90}{\footnotesize $\TR\,[\GEV\,]$}}
  \put(0.46,0.265){\rotatebox{-90}{\footnotesize $\sigma_{14\TEV}\;[\fb\,]$}}
  }
\end{picture}
\caption{Allowed points of the $(17+1)$-dimensional pMSSM scan in the plane spanned by
$\mG$ and $\TR$ (left panel) as well as $1\!-\mG^2/\mstau^2$ and $\TR$ (right panel). The
color encodes the total SUSY production cross section at the $14\TEV$ LHC\@.
}
\label{fig:grav4}
\end{figure}

Finally, we want to comment on the prospects of studying these scenarios 
at the upcoming long-term run of the LHC\@. Figure \ref{fig:grav4} shows the 
full SUSY cross section of the points that have passed all bounds discussed 
above. The points that are closest to the exclusion limit from the HSCP
search at $7$ and $8\TEV$ typically provide a SUSY cross section at the 
$14\TEV$ LHC run of $\sigma_{14\TEV}^{\text{SUSY}}\simeq100\fb$, corresponding
to the red points in figure \ref{fig:grav4}. 
Since the cross section can have a strong dependence on sectors that are rather
decoupled from the physics that constrain the reheating temperature---like 
the masses of the first generation squarks---the variation of the point color is relatively 
uncorrelated. However, we see that the uppermost stripe of the allowed band in the 
left panel does not contain points with very small cross sections due to the generically 
lighter gauginos for larger reheating temperatures. Many points in our scan with 
$\TR>10^9\GEV$ provide cross sections around $1\fb$ or higher.

Since the points with $\TR>10^9\GEV$ all feature the resonant configuration
$m_A\simeq2\mstau$, at the LHC the direct stau production via a resonant heavy 
Higgs in the $s$-channel will be an important production mechanism \cite{Lindert:2011td}. 
For this process the production near threshold is significantly enhanced and the velocity 
distribution of the staus peaks at rather low values $\beta\lesssim0.4$ \cite{Lindert:2011td}.
Such a signature is expected to be challenging for the current trigger settings at ATLAS 
and CMS and may require an extended buffering of the tracker data as pointed out in
\cite{Heisig:2012zq}. Further, providing rather slow staus, a noticeable amount of staus 
might be trapped inside the detector and eventually decay into the gravitino and a tau. 
Potentially this enables the determination of the stau life-time
\cite{Asai:2009ka,Pinfold:2010aq,Graham:2011ah}. This is particularly interesting 
regarding the fact that a possible determination of the gravitino mass from the detection
of the tau requires the tau energy,
\beq
E_\tau\simeq\frac{\mstau}{2}\left(1-\frac{\mG^2}{\mstau^2}\right)\,,
\eeq
to deviate significantly from $\mstau/2$, i.e., $\mG^2\;\;/\!\!\!\!\!\!\!\ll\mstau^2$ \cite{Martyn:2006as}.
In the right panel of figure~\ref{fig:grav4} we show the allowed points in the plane spanned by 
$1-\mG^2/\mstau^2$ and $\TR$. Points with large $\TR$ tend to have values $1-\mG^2/\mstau^2$
that deviate significantly from one.
Therefore, the prospects of testing supergravity by the simultaneous measurement of 
$\mstau$, $\tau_{\stau_1}$ and $\mG$ \cite{Buchmuller:2004rq,Feng:2004gn}---allowing 
the verification of \eqref{eq:taustau}---are significantly better in these scenarios, featuring 
large gravitino masses, than in scenarios with smaller gravitino masses and therefore 
smaller $\TR$.

\section{Conclusions} \label{sec:Conclusion}

We worked out the interplay between constraints on the SUSY parameter space
and the highest possible reheating temperatures in a gravitino-stau scenario.
We performed a Monte Carlo scan over a $(17+1)$-dimensional parameter space.
By demanding that the gravitino abundance matches the measured DM
abundance we computed the required reheating temperature for each scan 
point taking into account the thermal and non-thermal production of gravitinos.
Both quantities depend non-trivially on the MSSM spectrum parameters. We derived 
the cosmological viability from the application of bounds from BBN and the diffuse 
gamma ray spectrum. According to the strong constraints imposed for large stau 
life-times, $\tau_{\stau_1}\gtrsim10^7\SEC$, from photodissociation processes
causing an overproduction of ${}^3$He, we do not encounter allowed points with 
stau life-times larger than $10^7\SEC$. 

We found valid points with a reheating temperature high enough to allow for thermal
leptogenesis, $\TR\gtrsim10^9\GEV$. These points are consistent with BBN bounds, 
flavor and precision bounds, theoretical bounds from vacuum stability, bounds from 
the HSCP searches at the 7 and 8\,TeV LHC as well as bounds from the MSSM Higgs 
searches and the requirement of providing a Higgs around 125\,GeV.
All these points lie in the resonant region, $m_A\simeq2\mstau$. In this region
annihilation dominantly takes place via the exchange of an $s$-channel heavy Higgs. 
For most of these points stau-pair annihilation is the dominant channel. 
However, we also found points where pair-annihilation of co-annihilating stops or 
EWinos is dominant. Most of the points with $\TR\gtrsim10^9\GEV$ have exceptionally 
low stau yields $10^{-16}<Y<10^{-14}$. Further, the separation in the mass between 
the stau and the gauginos tends to be small especially for points with larger yields. 
This tendency is most pronounced for $M_2$. This is due to the fact that the 
abundance of thermally produced gravitinos is approximately proportional to 
$g_i^2M_i^2$ evaluated at the scale $\TR$. Compared to $M_2$ the slower running 
of $M_3$ up to the scale $\TR$ over-compensates the effect of the larger coupling for 
the strong interaction.

For most of the points with $\TR>10^9\GEV$ the dominant production mode at the 
$14\TEV$ LHC will be the production of EWinos or gluinos being relatively close in 
mass to the stau. However, due to the resonant configuration, $m_A\simeq2\mstau$, 
resonant stau production via the $s$-channel heavy Higgs will be an important contribution. 
This leads to the signature of extremely slowly moving heavy stable charged sparticles. 
For such a signature one would greatly benefit from an extended buffering of the tracker 
data in the LHC detectors increasing the trigger efficiencies for staus that arrive largely 
delayed in the muon chambers. Further, the signature can lead to a large amount of staus 
that are stopped in the detectors. This could provide the intriguing possibility of measuring 
the stau life-time. Moreover, especially for a heavy gravitino as required in order to obtain 
a high reheating temperature the determination of the gravitino mass might be possible 
from the measurement of the energy of the tau that is produced in the decay of the stopped 
stau. The combination of a variety of bounds on the low-scale SUSY parameters has 
pointed us to a very interesting corner in parameter space that should be looked at in the
upcoming LHC run.

\subsection*{Acknowledgements}

I would like to thank Torsten Bringmann, J\"orn Kersten, Boris Panes and 
Tania Robens for very helpful discussions.
This work was supported by the German Research Foundation (DFG) via the
Junior Research Group `SUSY Phenomenology' within the Collaborative
Research Center 676 `Particles, Strings and the Early Universe'.

\addcontentsline{toc}{chapter}{References}
\bibliographystyle{../../utphys}
\bibliography{../../staus}

\end{document}